\documentstyle[12pt,aaspp4,psfig]{article}

\lefthead{left}
\righthead{right}

\newcommand{\gtilde}
 {~ \raisebox{-1ex}{$\stackrel{\textstyle >}{\sim}$} ~}
\newcommand{\ltilde}
 {~ \raisebox{-1ex}{$\stackrel{\textstyle <}{\sim}$} ~}

\begin{document}

\vspace*{-1cm}
\noindent
\hspace*{12cm}
UTAP-264/97 \\

\title{Cosmological Gamma-Ray Bursts and Evolution of Galaxies}

\author{Tomonori Totani}
\affil{Department of Physics, School of Science,
The University of Tokyo, Tokyo 113,
Japan \\ E-mail: totani@utaphp2.phys.s.u-tokyo.ac.jp}

\begin{center}
To Appear in ApJ Letters \\
(Received 1997 May 15; Accepted 1997 July 2)
\end{center}

\begin{abstract}
Evolution of the rate density of cosmological gamma-ray bursts (GRBs) is
calculated and compared to the BATSE brightness
distribution in the context of binary neutron-star mergers as the
source of GRBs, taking account of the realistic star formation history 
in the universe and evolution of compact binary systems. 
We tried two models of the evolution of cosmic star formation rate (SFR):
one is based on recent observations of SFRs at high
redshifts, while the other is based on a galaxy evolution model
of stellar population synthesis that reproduces the
present-day colors of galaxies. It is shown that the binary merger
scenario of GRBs naturally results in the comoving rate-density
evolution of $\propto (1+z)^{2-2.5}$ up to $z \sim $ 1, that has been
suggested independently from the compatibility between the number-brightness
distribution and duration-brightness correlation.
If the cosmic SFR has its peak at $z \sim $ 1--2 as suggested by
recent observations, 
the effective power-index of GRB photon spectrum, $\alpha \gtilde 1.5$ is
favored, that is softer than
the recent observational determination of $\alpha = 1.1 \pm 0.3$.
However, high redshift starbursts ($z \gtilde 5$) in elliptical galaxies,
that have not yet been detected, can 
alleviate this discrepancy. The redshift of GRB970508 is likely
about 2, just below the upper limit that is recently determined, and
the absorption system at $z = 0.835$ seems not to be the site of the GRB.
\end{abstract}


\section{Introduction}
It is well known that spatial distribution of 
the classical gamma-ray bursts (GRBs) detected by
the Burst and Transient Source Experiment (BATSE; Meegan et al. 1996)
is isotropic with high precision but the number of weak 
bursts is significantly deficient
compared with the Euclidean distribution (Meegan et al. 1992),
suggesting that GRBs are located at cosmological distances. Furthermore,
recent discovery of
an optical transient source embedded in an extended object for GRB970228
(e.g., van Paradijs et al. 1997; Sahu et al. 1997),
or metal absorption lines at $z = 0.835$ for GRB970508 (Metzger et al. 1997)
finally confirmed the cosmological origin of GRBs.
The observed log$N$-log$P$
distribution, where $N$ is the observed number of GRBs with peak
photon flux larger than $P$ [cm$^{-2}$s$^{-1}$], 
agrees well with a cosmological
distribution (Mao \& Paczy\'{n}ski 1992; Piran 1992; Dermer 1992) 
if the faintest bursts
are located at redshift of $z \sim 1$ and the comoving GRB rate
density is constant with time. The effect of 
possible GRB rate evolution has also been discussed in a number of papers
(see, e.g., Cohen \& Piran 1995; Rutledge, Hui, \& Lewin 1995)
using some analytic forms of
GRB rate evolution, e.g.,  $R_{\rm GRB} \propto (1+z)^\beta$.

The best candidate for the cosmological origin of GRBs is
widely considered to be mergers of binary neutron stars
(see, e.g., Narayan, Paczy\'{n}ski, and Piran 1992, and references
therein). Lipunov et al. (1995) analyzed the BATSE log$N$-log$P$ 
distribution based on this scenario, taking account of
the evolution of binary merger rate calculated by detailed
Monte-Carlo simulations of the binary system evolution. On the other hand, the
cosmic evolution of star formation rate (SFR) also plays an important
role in predicting the cosmic evolution of compact binary merger rate.
Lipunov et al. classified all galaxies into elliptical and spiral
galaxies, and assumed initial star bursts for ellipticals and
constant SFR for spirals. 
Recent progress of observation of high redshift galaxies, however,
gives more detailed information on the cosmic star formation history
(Lilly et al. 1996; Madau et al. 1996). The Canada-France Redshift
Survey (CFRS) revealed a marked evolution of 2800 {\AA} luminosity
density, that is considered to be a star formation
indicator, as ${\cal L}_{2800} \propto 
(1+z)^{3.9 \pm 0.75}$ to $z \sim 1$ (for $\Omega_0 = 1$, Lilly et al. 1996).
The constant SFR approximation in spiral galaxies is therefore no longer
justified even at $z < 1$.

Because the star formation history in a galaxy is strongly 
correlated to its present-day colors or spectra, that are different
among the morphological types of galaxies, it is also possible
to construct a model of cosmic star formation history based on
the spectrum of local galaxies by using
galaxy evolution models of stellar population synthesis (Totani,
Sato, \& Yoshii 1996; Totani, Yoshii, \& Sato 1997). In this Letter,
we analyze the BATSE log$N$-log$P$ distribution 
based on the realistic models of the cosmic
star formation history in the context of the compact binary mergers
as the source of GRBs, taking account of the evolution of
compact binaries due to the gravitational wave radiation.
We use two
models of the star formation history: one is based on the recent observations
of high redshift SFRs, and the other is a theoretical model based on
the galaxy evolution models of stellar population synthesis (Arimoto \&
Yoshii 1987; Arimoto, Yoshii, \& Takahara 1992, hereafter AYT).

\section{Models of Cosmic GRB Rate History}
The cosmic evolution of comoving merger-rate density 
of binary neutron stars ($R_{GRB}$)
as a function of cosmic time $t_c$ 
is given by a convolution of cosmic SFR density ($R_*$) and
probability distribution of the time from formation to
merger of a binary system ($P_m$), i.e., 
\begin{equation}
R_{\rm GRB}(t_c) \propto \int_{t_F}^{t_c}dt' R_*(t') P_m(t_c - t') \ ,
\label{eq:grb-rate}
\end{equation}
where $t_F$ is the formation
epoch of galaxies. We have assumed here
the binary formation rate is proportional to the star formation rate.
Because the massive stars in high-mass binary systems
evolve into double neutron stars
with much shorter time scale than the 
typical time scale of galaxy evolution or the
Hubble time (1--10 Gyrs), the form of $P_m$ is essentially
determined by the initial distribution of the separation between
two neutron stars. Since a compact binary with separation $a$ will
merge in a time $t \propto a^4$ by gravitational wave radiation,
$P_m(t)$ can be written as
\begin{equation}
P_m(t) \propto \frac{dn}{dt} = \frac{dn}{da}
\frac{da}{dt} \propto t^{\gamma/4} t^{-3/4} \ ,
\end{equation}
where we assumed the initial separation distribution as
$dn/da \propto a^\gamma$. If we assume $\gamma = -1$,
as observed in the distribution of initial separation of normal
binaries (i.e., double main-sequence stars; Abt 1983), $P_m$ is
proportional to $t^{-1}$. Although
there is few observational information on $dn/da$ of compact binaries, the
dependence of $P_m(t)$ on the uncertain $\gamma$ is small 
because of the fourth-power dependence of merger time on $a$. 
There are some calculations of $P_m$
in a more sophisticated way using population synthesis models
of stellar binary systems
(Tutukov \& Yungelson 1994; Lipunov et al. 1995),
and their calculations
also show that $P_m$ is approximately described by the form $P_m \propto
t^{-1}$, with the lower cut-off of $t_l \sim$ 0.02 Gyr that corresponds to
initial separation of $\sim 1 R_\odot$. In the following 
analysis, we assume that $P_m(t) \propto t^{-1}$ when $t \geq t_l$= 0.02 Gyr,
while $P_m = 0$ when $0 \leq t < t_l$.

\begin{figure}
  \begin{center}
    \leavevmode\psfig{figure=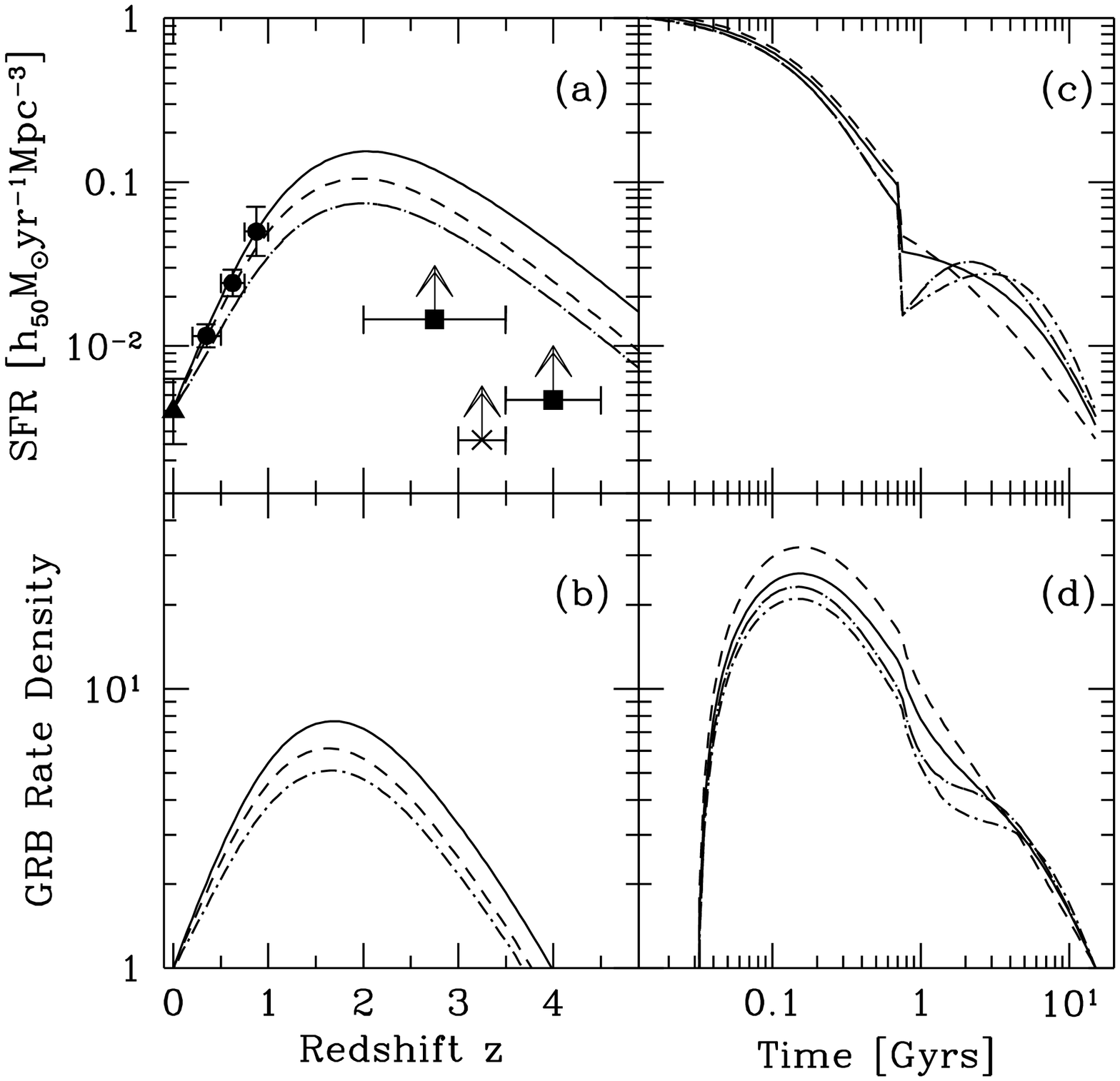,width=10cm}
  \end{center}
\caption{($a$) Data points are the recent measurements of 
comoving density of star formation rate (SFR) 
(Madau et al. 1996, for $\Omega_0 = 1$ universe), 
and the curves are the models of
SFR history based on these observations (see text). The cosmological
parameters of $(\Omega_0,
\lambda_0) = (1, 0)$, (0.2, 0) and (0.2, 0.8) are used for 
the solid, dashed, and dot-dashed lines, respectively.
($b$) Models of the comoving rate-density
evolution of GRBS calculated from the SFR history shown in ($a$). The curves
are normalized at $z$ = 0.
($c$) SFR history
calculated from the galaxy evolution model of population synthesis
as a function of time from galaxy formation.
Four curves represent the models of spiral galaxies: S1 (solid),
S2 (dashed), I1 (dot-short-dashed) and I2 (dot-long-dashed) (see text
for detail). ($d$) GRB rate evolution calculated from the SFR history shown
in ($c$). The curves are normalized at 12 Gyr.}
\label{fig:grb-rate}
\end{figure}

In the upper panels of Fig. \ref{fig:grb-rate}, we show the models of
cosmic star formation history used in this Letter. 
Fig. \ref{fig:grb-rate} (a) shows the 
``observational'' model of comoving SFR density
evolution, with recent observational
estimates of the SFR density at high redshifts (Madau et al. 1996).
In the range of $z$ = 0--1, the model is adjusted to the data
of the CFRS (Lilly et al. 1996), 
and the model SFR beyond $z=1$ is {\it not} just a simple
extrapolation, but it is based on
the evolution of neutral hydrogen gas in damped Ly$\alpha$
systems seen in quasar spectra that implies a peak of cosmic SFR 
around $z \sim$ 1--2 (Pei \& Fall 1995). 
The observed SFR density, that is inferred from the 
luminosity density of galaxies,  generally 
depends on the cosmological parameters ($\Omega_0$ and
$\lambda_0$), and we assume the same dependence
of the SFR evolution on $\Omega_0$ and $\lambda_0$ as that of luminosity
density. Fig. \ref{fig:grb-rate} (b) shows the comoving rate-density
of GRBs calculated by Eq. (\ref{eq:grb-rate}) 
with the SFR models shown in Fig. \ref{fig:grb-rate} (a),
for which we normalize the present GRB rate to the unity. 
(The absolute values of GRB rates are determined by the fit to
the BATSE data.) An important result is that
the GRB rate evolves rapidly from $z = 0$ to $z \sim 1$, and
has its peak at $z \sim$ 1--2.
If we parametrize the evolution of GRB rate
as $(1+z)^\beta$, these calculations imply that $\beta$ lies in a range of
2--2.5 at $z \ltilde 1$,
depending on the cosmological parameters.

Fig. \ref{fig:grb-rate} (c) shows the
SFR evolution calculated from the population synthesis
model of galaxy evolution. For details of the calculation,
see Totani et al. (1996; 1997). The SFR evolution is given
as a function of time from the galaxy formation, and four different
curves correspond to the four variations of the galaxy evolution model
for spiral galaxies: S1, S2, I1, and I2.  
The symbol `S' refers to the simple, closed-box model, while `I' to the 
infall model that allows for material infall into the disk region.  
The number attached to `S' and `I' is the adopted value of power index 
in the Schmidt law of star formation (for details see AYT). 
These four models are used to
assess the uncertainties in the star formation history of the 
galaxy evolution model. The model for elliptical galaxies is a 
so-called galactic wind model, in which stars are formed by the
initial star bursts and there is no star formation after the galactic 
wind that occurs about 1 Gyr after the galaxy formation. The SFR
in the universe is dominated by ellipticals during the first 1 Gyr
after the formation. 
The GRB rate evolutions calculated from these models of SFR evolution are
shown in Fig. \ref{fig:grb-rate} (d).

The observational SFR evolution [Fig. \ref{fig:grb-rate} (a)] and 
that of the galaxy evolution model [Fig. \ref{fig:grb-rate} (c)] 
are consistent with each other at 
$z \ltilde 1$ where SFR is
dominated by spiral galaxies, provided that there is non-vanishing
cosmological constant, $\lambda_0$ (Totani, Yoshii, \& Sato 1997).
However, beyond $z \sim$ 2--3 or during $\sim 1$ Gyr after the galaxy
formation, these two models give completely different behaviors
because the initial star bursts of giant elliptical galaxies are 
included in the galaxy evolution model while the signature of 
such bursts has not yet been detected. Some interpretations for this
problem are proposed,
but here we take the picture that giant elliptical
galaxies formed at $z \gtilde$ 5 and no star formation occurs
at observed redshifts (Totani, Yoshii, \& Sato 1997; Maoz 1997). 
We will discuss later the possibility that the GRB 
peak flux distribution can provide some information on the formation
epoch of elliptical galaxies.

\section{Results: Comparison of GRB Rate Evolution and the BATSE data}
For the comparison of the predicted GRB rate evolution and
the BATSE data, we use the number versus
peak flux distribution of the BATSE 3B catalog (Meegan et al. 1996)
measured in the energy range of 50--300 keV
by 1024 msec time scale. We set the analysis threshold of $P_{th} =
0.4$ [photons cm$^{-2}$s$^{-1}$] above which the detection efficiency
is almost 100 \%, and there are 665 GRBs above this threshold.
We assume neither dispersion nor evolution of the intrinsic luminosity
of GRBs (standard candle approximation). The detection rate of GRBs ($N$)
whose peak flux is larger than $P$ is calculated as follows:
\begin{equation}
N(>P) = \int_0^{z(P)} \frac{dV}{dz} \frac{R_{\rm GRB}(z)}{(1+z)} dz \ ,
\end{equation}
where $dV/dz$ is the comoving volume element per unit $z$, and $z(P)$ is 
a redshift that corresponds to the peak flux $P$. The factor of 
$1/(1+z)$ is included to account for the time dilation of the interval
between detected bursts. The value of $z(P)$
is given as the solution of the following equation,
\begin{equation}
\frac{P}{P_{th}} = \left(\frac{1+z}{1+z_{th}}\right)^{2-\alpha}
\left(\frac{d_L(z_{th})}{d_L(z)}\right)^2  \ ,
\end{equation}
where $d_L$ is the luminosity distance, $z_{th}$
the redshift corresponding to $P_{th}$ (hereafter $z_{0.4}$),
and $\alpha$ the effective power-index of photon number spectrum of
GRBs ($dN/dE_\gamma \propto E_\gamma^{- \alpha}$).
A recent investigation by Mallozzi, Pendleton, \& Paciesas (1996, 
hereafter MPP96) showed that the range of $\alpha$ appropriate for the
log$N$-log$P$ analysis is 1.1 $\pm$ 0.3.
However, the dispersion of spectral index is considerably large within
individual bursts, and we will try a wide range of $\alpha$ in the following,
considering the MPP96 range to be feasible.

\begin{figure}
  \begin{center}
    \leavevmode\psfig{figure=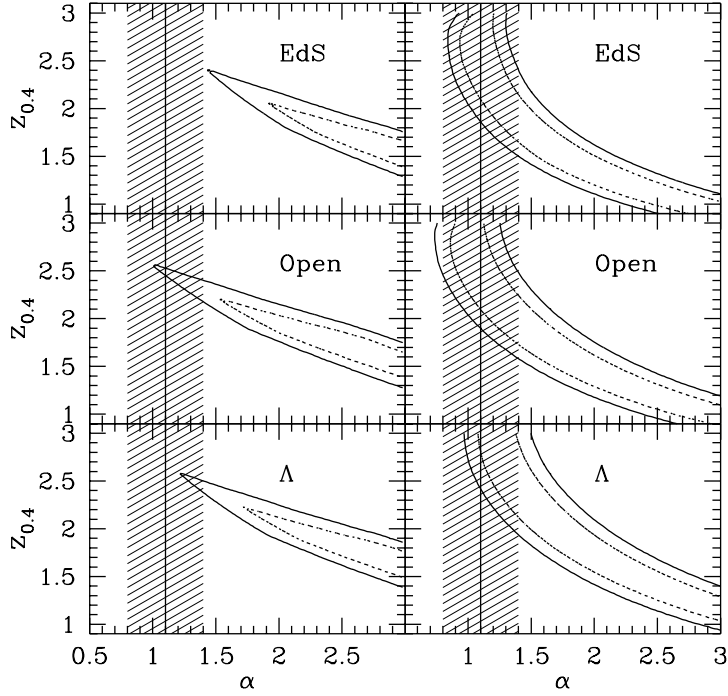,width=10cm}
  \end{center}
\caption{Allowed regions for the
effective power-index of GRB spectra $(\alpha)$ and the redshift
corresponding to the BATSE peak flux of 0.4 [cm$^{-2}$s$^{-1}$]
($z_{0.4}$) obtained from the log$N$-log$P$ analysis of the BATSE data.
The dotted lines are for 68 \% C.L. regions, while the 
solid lines for 95\% C.L.
The observational SFR-evolution model is used in the 
left panels, while the SFR of
galaxy evolution model (I1) is used for right panels. The cosmological
parameters of $(\Omega_0, \lambda_0) = (1, 0), (0.2, 0)$ and (0.2,
0.8) are used for the top, middle, and bottom panels, respectively.
The shaded regions are the feasible range of $\alpha$ determined
by the BATSE data (Mallozzi, Pendleton, \& Paciesas 1996).
}
\label{fig:zmax-alpha}
\end{figure}

We use the Kolmogorov-Smirnov test for comparison between the models
and the BATSE log$N$-log$P$ distribution. Fig. \ref{fig:zmax-alpha}
shows the allowed regions 
obtained by this test as contour maps of $z_{0.4}$ and
$\alpha$, for the three representative cosmological models.
The ``observational'' GRB rate model is used in the left panels,
while the GRB rate calculated from the galaxy evolution model (I1) in
the right panels. 
The formation redshift of galaxies is assumed to be $z_F = 5$ for all types
of galaxies and the effect of changing $z_F$ is small as long as the
redshift of the faintest bursts observed by BATSE is less than
$\sim 3$. We have chosen $H_0$ so as to
set the age of galaxies to a reasonable value (12 Gyrs).
A general trend is that the
allowed range for $z_{0.4}$ becomes smaller with increasing 
$\alpha$, because larger values of $\alpha$ (i.e., softer spectra)
make the cosmological
effect stronger, and $z_{0.4}$ has to become smaller to compensate
the cosmological effect.
The left panels of this figure show that the observational SFR history favors
$\alpha \gtilde 1.5$ because of the rapid decline
of $R_{GRB}(z)$ beyond $z \sim $2, and this is inconsistent with the
determination by MPP96, although the significance of this discrepancy
is difficult to assess due to the uncertainties in estimation of
$\alpha$ as well as in SFR observations. On the other hand, 
the results of the GRB rate derived from the galaxy evolution model
(right panels) show that the existence of high redshift starbursts
in elliptical galaxies alleviates the above discrepancy,
because the GRB rate continues to increase toward the past
to redshift larger than $\sim$ 2--3. There are allowed regions
for $z_{0.4} \sim$ 2--3 with the spectral index determined by MPP96.
The spiral models other than I1 give qualitatively similar results,
and $z_{0.4}$ varies by about $\ltilde$ 1 with changing models.

In Fig. \ref{fig:logNlogP} we show some examples of 
the predicted log$N$-log$P$
distribution (upper panel) and corresponding $z$-$P$ relation (lower panel)
for the two SFR-evolution models:
the galaxy evolution model (thin-solid, short- and long-dashed lines)
and the observational SFR model (dot-dashed line). The adopted values
of $\Omega_0$, $\lambda_0$, $z_{0.4}$, and $\alpha$ are shown in the figure.
All the four curves are consistent with the BATSE data (see Fig. 
\ref{fig:zmax-alpha}), and have the roughly Euclidean slope ($-3/2$) 
at $P \gtilde$ 10 cm$^{-2}$s$^{-1}$ that is consistent with the {\it Pioneer
Venus Orbiter} data (Fenimore et al. 1993). If it is the case
that elliptical galaxies formed before $z \sim 5$, the observed
number of GRBs will continue to increase with decreasing flux
below the BATSE detection limit.
On the other hand, if the cosmic SFR is actually peaked at $z
\sim$ 1--2, we will see rapid decline of the number of GRBs below the
BATSE limit (dot-dashed line). 
In any case, the redshift corresponding to a peak flux
increases rapidly with decreasing flux below the BATSE limit, and
future experiments with greater sensitivity will provide valuable
information on the formation epoch of galaxies, especially
for ellipticals.

\section{Discussion \& Conclusions}
\begin{figure}
  \begin{center}
    \leavevmode\psfig{figure=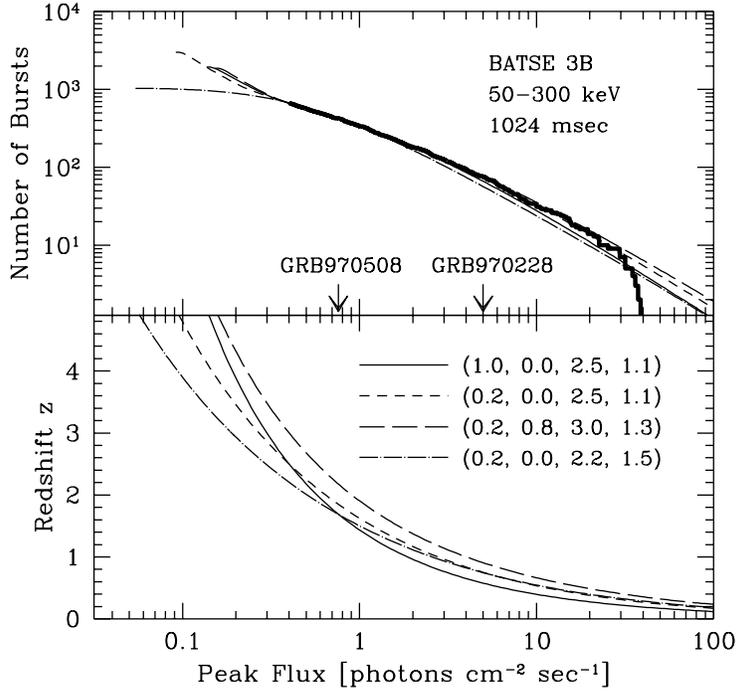,width=10cm}
  \end{center}
\caption{The number versus peak-flux distribution (upper panel) and
redshift versus peak-flux relation (lower panel).
Thick solid line is the observation by BATSE (Meegan et al. 1996).
Theoretical curves are calculated
with SFR of the galaxy evolution model (thin-solid, long- and
short-dashed lines)
and with the observational SFR-evolution model (dot-dashed line).
The line markings are the same for upper and lower panels.
The adopted values of $\Omega_0$, $\lambda_0$, $z_{0.4}$, and $\alpha$
are shown in the figure. The peak fluxes of GRB970228 and
GRB970508 are also indicated.}
\label{fig:logNlogP}
\end{figure}

Recent discovery of the metal absorption lines in the optical
counterpart associated with GRB970508 provides us an important
and independent constraint on the distance to the source of the GRB:
$0.835 \leq z < 2.1$ (Metzger et al. 1997). 
We calculate the peak fluxes in the BATSE range (50--300 keV) 
for GRB970508 (Kouveliotou et al. 1997), as well as for
GRB970228 (Costa et al. 1997; see also Piro et al. 1997 for
the conversion of BeppoSax count rate into photon flux),
assuming $\alpha = 1.1$. 
The fluxes are indicated in Fig. \ref{fig:logNlogP}. A striking
implication of our calculation is that the redshift of GRB970508
is likely $\sim$ 2, just below the upper limit of Metzger et al.
Although the possible dispersion in GRB luminosity function may
allow lower redshifts, it seems difficult to consider the absorption
system at $z = 0.835$ as the GRB source, as long as we assume that 
GRBs are produced by binary neutron-star mergers.

Some cosmological GRB models other than the compact binary mergers, e.g.,
failed Ib supernovae (Woosley 1993), 
predict the GRB rate evolution that is proportional to SFR. 
We analyzed the BATSE data
with the GRB rate proportional to the models of the cosmic SFR history
used in this Letter.
The GRB rate evolution at $z$ = 0--1
becomes steeper than the case of binary
mergers because of the lack of time lag during the spiral-in of
compact binaries by gravitational wave radiation, 
and this requires larger $z_{0.4}$ or $\alpha$
to compensate the steeper evolution of $R_{GRB}$. 
Therefore the allowed values for $\alpha$ become even softer,
apart from the determination of MPP96. Especially, with the
observational SFR evolution, we found no acceptable fit with $\alpha
< 1.5$ with 95 \% C.L., and it can be concluded that the compact
binary scenario is more favorable than GRB models that predict
a GRB rate proportional to SFR.

We have not considered the normalization of $R_{GRB}$ from the SFR
models, because the normalization does not affect the log$N$-log$P$
analysis. However the comparison of the model $R_{GRB}$ and
observed number of GRBs gives an important consistency check.
The fit of our typical $R_{GRB}$ models to the BATSE data implies
that the GRB rate at $z$ = 0 is $\sim 7 \times
10^{-9}h^3$ [yr$^{-1}$Mpc$^{-3}$] where $h = H_0$/(100 km/s/Mpc). 
On the other hand, assuming
the Salpeter IMF, flat distribution of the mass ratio
of binary systems, and binary formation rate equal to
SFR, we can estimate the binary merger rate expected from
the cosmic SFR history, that becomes about $6 \times 10^{-6}h^2$
[yr$^{-1}$Mpc$^{-3}$] at $z = 0$. The expected rate is therefore 
$\sim 10^3$ times larger than the observation of BATSE, and
the beaming of GRBs with $d\Omega \sim 4 \pi 10^{-3}$ may be required.
This result is consistent with the previous estimate of Lipunov et al.
(1995), although it should be noted that the above estimate has
large uncertainties in SFR itself or conversion of SFR into merger rate.

We have shown in this Letter, for the first time, 
that the binary neutron-star merger scenario
of the cosmological GRBs naturally results in the comoving rate-density 
evolution of GRBs
roughly proportional to $(1+z)^{2-2.5}$ at low redshifts ($z \ltilde$ 1).
It should be noted that such evolution has been suggested independently
from the compatibility between the log$N$-log$P$ analysis and 
time dilation analysis
(Horack, Emslie, \& Hartmann 1995; Horack, Mallozzi, \& Koshut 1996;
M\'{e}sz\'{a}ros \& M\'{e}sz\'{a}ros 1996). In fact, the time dilation 
factor, $(1+z_{\rm dim})/(1+z_{\rm bright})$, for the bright and 
dim+dimmest bursts defined by Norris et al. (1995) is
2.0--2.3 in the four curves depicted in the lower panel of
Fig. \ref{fig:logNlogP}, in nice agreement with the result of Norris
et al. (1995).

The author is grateful to an anonymous referee for valuable comments.
He also thanks A. M\'{e}sz\'{a}ros for useful comments.
This work has been supported in part by the Grant-in-Aid for the 
Scientific Research Fund (3730) of the Ministry of Education, Science,
and Culture in Japan.

\end{document}